\documentclass[useAMS]{mn2e}
\usepackage{graphicx}
\usepackage{subfig}
\usepackage{amssymb}
\usepackage{natbib}
\usepackage{amsmath}
\usepackage{multirow} 
\usepackage{booktabs,threeparttable}
\usepackage{color}
\usepackage[colorlinks,citecolor=black]{hyperref}
\usepackage{graphicx,url,times}
\usepackage{array}
\usepackage{gensymb}
\usepackage{hyperref}
\usepackage{placeins}
\usepackage{dcolumn}
\usepackage[varg]{newtxmath}
\DeclareGraphicsExtensions{.ps,.pdf,.png}
\makeatletter
\def\fps@figure{htbp}
\makeatother

\definecolor{grey}{rgb}{0.4,0.6,0.6}
\newcolumntype{L}{D{.}{.}{2,1}}

\vspace{-1cm}

\newcolumntype{P}[1]{>{\raggedleft\arraybackslash}p{#1}}

\begin{document}
\title[Galaxy Cluster Mass Reconstruction]{Galaxy Cluster Mass Reconstruction Project: \\
III. The impact of dynamical substructure on cluster mass estimates}
\author[Old et al.]{L. Old$^{1}$\thanks{E-mail:
old@astro.utoronto.ca},
R. Wojtak$^{2,3,4}$, F. R. Pearce$^{5}$, M. E. Gray$^{5}$, G. A. Mamon$^{6}$, C. Sif\'on$^{7,8}$,  E. Tempel$^{9,10}$, \newauthor A. Biviano$^{11}$, H. K. C. Yee$^{1}$, R. de Carvalho$^{12}$, V. M\"uller $^{9}$, T. Sepp$^{10}$,  R. A. Skibba$^{13,14}$, \newauthor D. Croton$^{15}$,  S. P. Bamford$^{5}$,  C. Power$^{16,17}$, A. von der Linden$^{18}$, A. Saro$^{19}$ \\
\\
\noindent Affiliations are listed at the end of the paper
\date{Accepted ??. Received ??; in original form ??}}
\pagerange{\pageref{firstpage}--\pageref{lastpage}} \pubyear{2017}
\maketitle
\label{firstpage}
\begin{abstract}
\noindent With the advent of wide-field cosmological surveys, we are approaching samples of hundreds of thousands of galaxy clusters. While such large numbers will help reduce statistical uncertainties, the control of systematics in cluster masses becomes ever more crucial. Here we examine the effects of an important source of systematic uncertainty in galaxy-based cluster mass estimation techniques: the presence of significant dynamical substructure. Dynamical substructure manifests as dynamically distinct subgroups in phase-space, indicating an ‘unrelaxed’ state.  This issue affects around a quarter of clusters in a generally selected sample. We employ a set of mock clusters whose masses have been measured homogeneously with commonly-used galaxy-based mass estimation techniques (kinematic, richness, caustic, radial methods).  We use these to study how the relation between observationally estimated and true cluster mass depends on the presence of substructure, as identified by various popular diagnostics. We find that the scatter for an ensemble of clusters does not increase dramatically for clusters with dynamical substructure. However, we find a systematic bias for all methods, such that clusters with significant substructure have higher measured masses than their relaxed counterparts. This bias depends on cluster mass: the most massive clusters are largely unaffected by the presence of significant substructure, but masses are significantly overestimated for lower mass clusters, by $\sim10\%$ at $10^{14}$ and $\ga20\%$ for $\la10^{13.5}$. The use of cluster samples with different levels of substructure can therefore bias certain cosmological parameters up to a level comparable to the typical uncertainties in current cosmological studies.
\end{abstract}


\section{Introduction}
Galaxy clusters are massive, rare objects which form from high peaks in the underlying density field and whose population characteristics are sensitive to the expansion history of the Universe and the growth rate of structure. Statistical studies of the galaxy cluster population are therefore powerful tools across various fields including cosmology (see \citealt{2005RvMP...77..207V}; \citealt{2011ARA&A..49..409A} for a review, \citealt{2012ApJ...745...16T}), galaxy evolution (e.g., \citealt{Dressler:1980wq}; \citealt{1999ApJ...527...54B}; \citealt{2003MNRAS.346..601G};
\citealt{2005ApJ...623..721P}; \citealt{2012ApJ...757....4P}) and large scale structure (e.g., \citealt{1988ARA&A..26..631B}; \citealt{2001AJ....122.2222E})\\
\indent We are entering an exciting time for cluster cosmology with ongoing surveys such as The Dark Energy Survey (\citealt{2005astro.ph.10346T}), the Kilo-Degree Survey (\citealt{2015A&A...582A..62D}), WFIRST (\citealt{2015arXiv150303757S}), the South Pole Telescope Sunyaev Zel’dovich survey (\citealt{2016ApJ...832...95D}), the Atacama Cosmology Telescope (\citealt{2011ApJ...732...44S}), the Hyper Suprime Cam survey (\citealt{2017arXiv170405858A}), and upcoming surveys such as Euclid (\citealt{2013LRR....16....6A}), eROSITA (\citealt{2012MNRAS.422...44P}) and LSST (\citealt{2009arXiv0912.0201L}). \\
\indent With the production of these wide-field surveys across a variety of wavelengths, we are moving into an era where samples of $10^{6}$ galaxy clusters will be available. These large samples enable the reduction of statistical uncertainties, however, it is clear that systematic uncertainties often dominate the statistical uncertainties in cluster mass estimation (as highlighted in \citealt{2013ApJ...763..147B}; \citealt{2013JCAP...07..008H}; \citealt{2016A&A...594A..24P}), and the need to control for these systematic uncertainties is even more crucial for cluster cosmology studies.\\
\indent One such source of systematic uncertainty in cluster mass estimation techniques in particular, is the presence of dynamically young clusters with significant dynamical substructure. Cluster dynamical substructure is characterised as the presence of dynamically distinct subgroups within galaxy clusters. In the cluster galaxy distribution substructure typically manifests itself in the form of asymmetrical velocity distributions and distinct subgroups in phase-space of clusters. The presence of significant substructure is an indication that a cluster is not in virial equilibrium or in a `relaxed' state, either because of a recent cluster-cluster merger, or significant growth of the cluster via infalling groups.\\
\indent There have been numerous studies since the 1980s probing the frequency of dynamical substructure in cluster samples (e.g., (\citealt{1982PASP...94..421G}; \citealt{1988AJ.....95..985D}; \citealt{1991A&A...246..301R}; \citealt{1994AJ....107.1637B}; \citealt{1994ApJ...423..539E}; \citealt{1995ApJ...451L...5W}; \citealt{1999A&A...343..733S}; \citealt{2004MNRAS.352..605B}; \citealt{2009ApJ...693..901O}; \citealt{2010A&A...521A..28A},  \citealt{2012MNRAS.420.2480Z}; \citealt{2012A&A...540A.123E}; \citealt{2012MNRAS.421.3594H}; \citealt{2012MNRAS.419.1017C},  \citealt{2017MNRAS.468.1824O}). Many of these works also explored whether measured global properties of clusters differ for clusters in their samples with significant substructure compared to more relaxed clusters. While some works have found that the measured global properties of clusters do differ in samples of clusters that have significant dynamical substructure (e.g., \citealt{1982PASP...94..421G}; \citealt{1994ApJ...423..539E}; \citealt{1995ApJ...451L...5W}; \citealt{1997ApJ...482...41G}; \citealt{2006ApJ...648..209L}; \citealt{2006A&A...456...23B}; \citealt{2012MNRAS.421.3594H}), other works do not find any obvious difference in cluster measures for complex clusters (e.g., \citealt{1993ApJ...411L..13B}; \citealt{1997ASPC..117..505F}; \citealt{2010A&A...521A..28A}). The discordance in the conclusions are likely due to small galaxy cluster samples and the method employed to characterise dynamical substructure.\\
\indent While these works focus on comparing measured global cluster properties for highly substructured and non-substructured clusters, in this study, we focus on deducing whether cluster mass estimation techniques themselves are affected by the presence of significant dynamical substructure, as opposed to differences in global parameters of these two cluster populations.\\
\indent One approach to examine whether cluster mass estimation techniques themselves are affected by the presence of significant dynamical substructure is to compare galaxy-based reconstructed mass estimates with reconstructed mass estimates computed using other mass proxies e.g, X-ray, lensing, SZ-based mass estimates. An example of this multi-wavelength comparison is in \citet{2006ApJ...648..209L}, where optical richness and X-ray luminosity relations for a sample of several hundred clusters are examined. The authors find that the exclusion of clusters with substructure does not improve the correlation between X-ray luminosity and richness, but does improve the relation between X-ray temperature and optical parameters. More recently, \citet{2013ApJ...772...25S} hints that disturbed systems may bias the relation between SZE-velocity dispersion cluster mass, however, they state the need for larger samples of clusters to confirm this.\\
\indent The second approach to deduce whether cluster mass estimation techniques themselves are affected by the presence of significant dynamical substructure, is to use mock data where the underlying halo mass is known, and global cluster properties including mass and relaxation state are measured in an observational manner. For example, \citet{1996ApJS..104....1P}, use N-body simulations of galaxy cluster mergers and find that virial masses are overestimated by up to a factor of 2 for clusters undergoing mergers, a conclusion similar to that of \citet{1990A&A...237..319P}.\\
\indent The main assumption required in this approach are that the simulated galaxy clusters deemed highly substructured by an observational substructure tests are indeed similar to clusters in the real Universe that would be deemed highly substructured by dynamical substructure tests. This assumption is reasonable in the case where the properties of galaxies in the simulated clusters used are taken directly from the underlying N-body dark matter simulation, where phase-space properties have primarily evolved over time due to the influence of gravity. To first order, these simulated phase-space properties are indeed comparable to galaxy phase-space properties in the Universe.\\
\indent To understand the consequence of including dynamically disturbed galaxy clusters in cluster cosmology samples, we look to examine the following questions: does the presence of significant dynamical substructure impact commonly used galaxy-based mass estimation techniques? 
Would scaling relations between multi-wavelength mass estimation techniques differ for highly substructured and non-substructured clusters? And finally, should dynamically young clusters be excluded from future cluster cosmology samples?\\
\indent In this work, we explore these critical questions, presenting the first extensive, homogenous study of the impact of dynamical substructure on galaxy-based cluster mass estimation techniques. We utilise part of the Galaxy Cluster Mass Reconstruction Project (GCMRP) dataset, where 25 different galaxy-based mass estimation techniques were tested using two mock galaxy catalogues to deduce how well these methods characterised global cluster properties such as mass (\citealt{2014MNRAS.441.1513O,2015MNRAS.449.1897O}), and how this mass depends on the accuracy of the selected members (Wojtak et al. in prep.)\\
\indent The article is organised as follows: we describe the mock galaxy catalogue in Section~2, and the mass reconstruction methods applied to this catalogue in Section~3. In Section~4, we provide details of our analysis before presenting the results on the effects of significant dynamical substructure on cluster mass estimation in Section~5. We end with a discussion of our results and conclusions in Section 6. Throughout the article we adopt a Lambda Cold Dark Matter ($\Lambda$CDM) cosmology with $\Omega_{\rm 0}=0.27$, $\Omega_{\rm \Lambda}=0.73$, $\sigma_{\rm
 8}=0.82$ and a Hubble constant of $H_{\rm 0} =
100\;h\,\rm{km\,s^{-1}}\,\rm{Mpc^{-1}}$ where $h=0.7$, although none of the conclusions depend strongly on these parameters.
\section{Data}
\label{sec:Sim Data}
\noindent For this study, we only use data from the GCMRP where the dynamical properties of the galaxies are taken directly from the underlying N-body dark matter subhaloes themselves, where the galaxies have retained the 'dynamical memory' of the merging history of the clusters. This strategy ensures a more direct comparison with that of the real Universe, where we assume the phase-space properties of galaxies have primarily evolved over time due to the influence of gravity. We take an observational approach in this study, measuring the dynamical state of our mock clusters using observational dynamical substructure tests. We describe the underlying dark matter simulation, light cone generation and model used to populate the dark matter simulation outputs with galaxies in the following subsections.
\subsection{Dark matter simulation}
\label{sec:DM data}
The underlying dark matter simulation we use is the Bolshoi dissipationless cosmological simulation which follows the evolution of $2048^{3}$ dark matter particles of mass $1.35 \times
10^{8}\,h^{-1} {\rm M_{\rm \odot}}$ from $z=80$ to $z=0$ within a box of side length $250\,h^{-1} {\rm Mpc}$ with a force resolution of the $1\,h^{-1}$ kpc (\citealt{2011ApJ...740..102K}). The simulation was run with the \textsc{ART} adaptive mesh refinement code following a flat $\Lambda$CDM cosmology with the following parameters: $\Omega_{\rm 0}=0.27$, $\Omega_{\rm \Lambda}=0.73$, $\sigma_{\rm
 8}=0.82$, $n=0.95$ and $h=0.70$. The halo catalogues are complete for haloes with circular velocity $V_{\rm circ}>50\,\rm{km\,s^{-1}}$ (corresponding to $M_{\rm 360\rho} \approx 1.5 \times
10^{10}\,h^{-1} {\rm M_{\rm \odot}}$, $\sim$110 particles).\\
\indent \textsc{ROCKSTAR}, a 6D FOF group-finder based on adaptive
hierarchical refinement, is used to identify dark matter haloes, substructure and tidal features (\citealt{2013ApJ...762..109B}).  \textsc{ROCKSTAR} identifies haloes and subhaloes using 6D (3D in spatial, and 3D in velocity) information which are joined into hierarchical merging trees that
describe in detail how structures grow as the universe evolves. As \textsc{ROCKSTAR} uses spatial and velocity information to identify dark matter structures, it does not suffer from (3D) projection effects that would potentially bias this study in incidences where two group centres were spatially aligned in the same snapshot. \textsc{ROCKSTAR} calculates the underlying halo masses by calculating the spherical overdensities according to a density threshold 200 times that of the critical density. We highlight that these overdensities are calculated using all the particles for all the substructure contained in a halo. This halo finder has been shown to recover halo properties with high accuracy and produces results consistent with those of other halo finders (\citealt{2011MNRAS.415.2293K}).  
\subsection{Light cone construction}
\label{sec:Light cone}
For this study, we use light cones produced by the Theoretical
Astrophysical Observatory
(TAO\footnote{\hyperref[https://tao.asvo.org.au/tao/]{https://tao.asvo.org.au/tao/}},
\citealt{2016ApJS..223....9B}), an online eResearch tool that provides access to semi-analytic galaxy formation models and N-body simulations. The light cone tool remaps the spatial and temporal positions of each galaxy in the simulation box onto a cone which subtends 60$^{\circ}$ by 60$^{\circ}$ on the sky, covering a redshift range of $ 0 < z < 0.15$. We specify a minimum $r$-band luminosity for the galaxies of $M_{r} = -19 + 5\log h$ for the catalogue.
\subsection{Semi-analytic model}
The model we use to form galaxies on the underlying dark matter data is the Semi-Analytic Galaxy Evolution (SAGE) galaxy formation model (\citealt{2016ApJS..222...22C}). As described in more detail in \citet{2015MNRAS.449.1897O}, this galaxy formation model is applied to the merger trees described in Section~\ref{sec:DM data}. In each tree and at each redshift, virialised dark matter haloes are assumed to attract pristine gas from the surrounding environment, from which galaxies form and evolve. The SAGE model is calibrated using various observations at $z=0$, namely the stellar mass function and SDSS-band luminosity functions, baryonic Tully-Fisher relation, metallicity-stellar mass relation and the black hole-bulge relation.\\
\indent The model includes various galaxy formation physics from reionisation of the inter-galactic medium at early times, the infall of this gas into haloes, radiative cooling of hot gas and the formation of cooling flows, star formation in the cold disk of galaxies and the resulting supernova feedback, black hole growth and active galactic nuclei (AGN) feedback through the `quasar' and `radio' epochs of AGN evolution, metal enrichment of the inter-galactic and intra-cluster medium from star formation, and galaxy morphology shaped through secular processes,  mergers and merger induced starbursts. Detailed comparisons of the model to observations at higher redshift can be found in \citet{2014ApJ...795..123L} and \citet{2016ApJS..222...22C}, though we note that our light cone spans only lower redshifts, as described in Section~\ref{sec:Light cone}.\\
\indent Importantly, each group identified by the halo finder \textsc{ROCKSTAR} has a `central' galaxy whose central position and velocity is determined by averaging the positions and velocities of the subset of halo particles. Each group also has a number of `satellite' galaxies  (cluster members) that maintain the positions and velocities of the subhaloes that merged with the parent halo.
\section{Mass Reconstruction Methods}
\label{sec:Mass Reconstruction Methods}
To determine the consequence of including dynamically disturbed galaxy clusters in cluster cosmology samples, we use a subset of the GCMRP dataset, where 23 commonly-used galaxy-based mass estimation techniques (kinematic, richness, caustic, radial methods), were tested in a blind manner on clusters from two mock galaxy catalogues. For this study, we use only results of galaxy-based techniques which were tested on mock clusters from the Semi-Analytic Model (SAM)-based dataset described in Section~2.3, where the dynamical properties of the galaxies are taken directly from the underlying N-body dark matter subhaloes themselves (unlike the HOD2 model used in \citealt{2015MNRAS.449.1897O}).\\
\indent The three general steps that galaxy-based techniques typically follow is first to locate the cluster overdensity, choose which galaxies are members of the cluster and finally use the properties of this membership to reconstruct cluster mass. In this study, we focus on the second and third steps of deducing membership and mass, as opposed to cluster finding. 
We summarise the type of data the methods require as input in Table~\ref{table:basic_method_characteristics} and the basic properties of all methods in Table~\ref{table:appendix_table_1} and  Table~\ref{table:appendix_table_2}, however, we refer the reader to studies \citet{2014MNRAS.441.1513O, 2015MNRAS.449.1897O} for more detail of the procedure of each cluster mass reconstruction technique. We note that the colour associated with each method in the figures and tables correspond to the main galaxy population property used to perform mass estimation richness (magenta), projected phase-space (black), radii (blue), velocity dispersion (red), or abundance matching (green).
\begin{table*} 
 \caption{Summary of the 23 cluster mass estimation
 methods. Listed is an acronym identifying the method, an
 indication of the main property used to undertake member galaxy selection
 and an indication of the method used to convert this membership
 list to a mass estimate. The type of observational data required as input for each method is listed in the fourth column. Note that acronyms denoted with an asterisk
 indicate that the method did not use our initial object
 target list but rather matched these locations at the end of their analysis. Please see Tables
 \ref{table:appendix_table_1} and \ref{table:appendix_table_2} in
 the appendix for more details on each method.} 
\begin{center} 
\tabcolsep 0.15cm
\begin{tabular}{l l l l l}
\hline
\multicolumn{1}{c}{Method}&Initial Galaxy Selection&Mass Estimation&Type of data required&Reference \\ \hline
PCN&phase-space&Richness&Spectroscopy&\citet{2015MNRAS.449.3082P}\\
PFN*&FOF&Richness&Spectroscopy&\citet{2015MNRAS.449.3082P}\\
NUM&phase-space&Richness&Spectroscopy&Mamon et al. (in prep.)\\
ESC&phase-space& phase-space&Spectroscopy&{\citet{2013ApJ...768L..32G}}\\ 
MPO&phase-space& phase-space&Multi-band photometry, spectroscopy&{\citet{2013MNRAS.429.3079M}}\\ 
MP1& phase-space& phase-space&Spectroscopy&{\citet{2013MNRAS.429.3079M}}\\
RW& phase-space& phase-space&Spectroscopy&{\citet{2009MNRAS.399..812W}}\\
TAR*&FOF& phase-space&Spectroscopy&{\citet{Tempel+14}} \\
PCO& phase-space&Radius&Spectroscopy&\citet{2015MNRAS.449.3082P}\\ 
PFO*&FOF& Radius& Spectroscopy&\citet{2015MNRAS.449.3082P}\\
PCR& phase-space&Radius&Spectroscopy&\citet{2015MNRAS.449.3082P}\\ 
PFR*&FOF&Radius&Spectroscopy&\citet{2015MNRAS.449.3082P}\\
MVM*&FOF&Abundance matching&Spectroscopy&{\citet{2012MNRAS.423.1583M}}\\
AS1&Red Sequence&Velocity dispersion&Spectroscopy&{\citet{2013ApJ...772...47S}}\\
AS2&Red Sequence&Velocity dispersion&Spectroscopy&{\citet{2013ApJ...772...47S}}\\
AvL& phase-space&Velocity dispersion&Spectroscopy&{\citet{2007MNRAS.379..867V}}\\ 
CLE& phase-space&Velocity dispersion&Spectroscopy&{\citet{2013MNRAS.429.3079M}}\\ 
CLN& phase-space&Velocity dispersion&Spectroscopy&{\citet{2013MNRAS.429.3079M}}\\
SG1& phase-space&Velocity dispersion&Spectroscopy&{\citet{2013ApJ...772...25S}}\\
SG2& phase-space&Velocity dispersion&Spectroscopy&{\citet{2013ApJ...772...25S}}\\
SG3& phase-space&Velocity dispersion&Spectroscopy&{\citet{2009MNRAS.392..135L}}\\
PCS& phase-space&Velocity dispersion&Spectroscopy&\citet{2015MNRAS.449.3082P}\\ 
PFS*&FOF&Velocity dispersion&Spectroscopy&\citet{2015MNRAS.449.3082P}\\ 
\hline 
\end{tabular}
\end{center}
\label{table:basic_method_characteristics}
\end{table*} 
\section{Dynamical substructure analysis}
The tools for detecting dynamical substructure, either solely using the cluster member velocity distribution (1D), the member positions (2D) or combining the velocity and positional information of the cluster (3D), have been extensively assessed for their robustness and reliability for both group sized systems and cluster sized systems (\citealt{1996ApJS..104....1P}; \citealt{2009ApJ...702.1199H}). These comprehensive works indicate that while applying a variety of 1D, 2D and 3D dynamical substructure tests is useful, the more reliable substructure tests are 3D tests which quantify the difference between local subgroups of galaxies within clusters to the global cluster properties such as the Dressler-Shectman (DS, 1988) test and the Kappa test (\citealt{1996ApJ...458..435C}). In this study, we apply these tests to our semi-analytic mock simulation data (where we again note that the mock galaxy properties are taken from the underlying N-body simulation dark matter subhaloes). A cluster is deemed as significantly dynamically substructured if either the DS test or the Kappa test detected substructure. We outline the procedure of these tests below.\\
\indent While these tests are found to be the more reliable techniques applied in the literature (see extensive evaluations in \citealt{1996ApJS..104....1P}; \citealt{2009ApJ...702.1199H}), there can be cases where clusters do indeed contain significant substructure undetected by these tests. For example, \citet{2010MNRAS.408.1818W}, use N-body simulations to test the correlation between a given dynamical substructure detection technique and time since last major merger of a cluster. They find that this correlation is dependent on viewing angle, especially in cases where the substructure is not well separated along the line of sight. Furthermore, \citet{2012MNRAS.421.3594H} find that the DS test in particular can be reliably applied to groups only with $N_{\rm gal}>20$ and where a high confidence level of $95\%$ or higher is used. Indeed, \citet{2012MNRAS.421.3594H} deduced that for groups with $10\leq N_{\rm gal}<20$, the DS test does not necessarily detect all substructures within a system, but the test can be used to determine a reliable lower limit on the amount of substructure.
\subsection{The Dressler-Shectman test}
\noindent The DS test aims to quantify the difference between local kinematics and global kinematics by selecting subgroups of cluster members and calculating the local velocity dispersion $\sigma_{\rm local}$ and velocity mean $\overline{\nu}_{\rm local}$. These local properties are compared with the global cluster velocity dispersion $\sigma_{\rm global}$ and cluster velocity mean $\overline{\nu}_{\rm global}$ by computing an $i$-th deviation $\delta_{i}$ for the $i$-th cluster member:
\begin{equation}
\delta_{\rm i}^{2}=\left(\frac{N_{\rm nn}+1}{\sigma_{\rm global}} \right )\, \\\left[(\overline{\nu}_{\rm local}-\overline{\nu}_{\rm global})^{2}+(\sigma_{\rm local}-\overline{\sigma}_{\rm global})^{2}\right].\
\end{equation}
We adopt a correction to the original DS test by replacing $N_{\rm nn} = 11$ with $N_{\rm nn} = \sqrt{ N_{\rm members}}$ as suggested for applying to groups and clusters with fewer members to enhance the sensitivity of the test to small-scale structures (\citealt{1986desd.book.....S} and \citealt{1998ApJ...498L...5Z}).
The deviations are then summed to give $\Delta$, the DS statistic\\
\begin{equation} 
\Delta= \displaystyle\sum\limits_{i} \delta_{i}.\end{equation}\\
Often referred to as the critical value for the cluster, the $\Delta$-statistic is used to compute a PTE for the presence of substructure by computing 10,000 Monte Carlo realisations, shuffling the member velocities amongst the positions. The PTE is used to test the null hypothesis that the cluster has no substructure, hence a small PTE $\leq 0.05$ indicates that the cluster has significant substructure.
\subsection{The Kappa test}
In addition to the DS test, we employ another 3D dynamical substructure test, the $\kappa$-test (\citealt{1996ApJ...458..435C}), which quantifies the difference between local substructures and global cluster phase-space properties using the Kolmogorov---Smirnov (KS) test. Similar to the DS-test, for each galaxy within the cluster, $N_{\rm nn} = \sqrt{ N_{\rm members}}$ nearest galaxies are selected and the velocity distribution of that local subgroup is compared to the parent distribution by measuring the maximum separation of the cumulative distribution functions $D_{\rm Obs}$. The negative log likelihood of producing a $D$-statistic greater than $D_{\rm Obs}$ is computed and summed for all $N$ galaxies in the cluster:\\
\begin{equation}
\kappa_{n} = \sum^{n}_{i=1}-[\rm{log}(P_{KS}(D_{\rm sim}>D_{Obs})].
\end{equation}
As for the DS test, the significance of the $\kappa_{n}$ statistic is computed by performing 10,000 Monte Carlo realisations, shuffling the member velocities amongst the positions to produce a Probability to Exceed (PTE). The PTE, $0 \leq \rm{PTE} \leq 1$, is used to test the null hypothesis that the cluster has no substructure, hence a small PTE $\leq 0.05$ indicates that the cluster has significant substructure. For clusters with $N_{\rm gal}$ $\geq$ 30, it is noted that the DS test is one of the most sensitive test for substructure detection \citep{1996ApJS..104....1P} and is reliable for clusters with $N_{\rm gal}\geq$ 20, provided that the PTE is 0.05 or 0.01 (\citealt{2000A&A...354..761K} and \citealt{2012MNRAS.421.3594H}). The test is also reliable to use as a lower limit for group sized systems with $N_{\rm gal}\geq$ 10.\\
\subsection{Mock cluster sample and analysis}
In this study, we apply the commonly used dynamical substructure DS and Kappa tests as described in the above sections on semi-analytic clusters whose galaxy properties are taken from the underlying N-body simulation dark matter subhaloes. A cluster is deemed as highly dynamically substructured if either of these tests detect substructure. As mentioned in Section~4, the dynamical substructure tests may not detect significant substructure in certain cases. This means that our sample of clusters that are deemed to be non-substructured, may have some level of contamination from substructured clusters. We first select all clusters with $N_{\rm gal} \geq20$ from the GCMRP cluster sample, leaving us with 943 clusters between $13.50 \leq $ log $(M_{\rm 200c, true}/M_{\sun})\leq 15.14$ and with a median mass of log $(M_{\rm 200c, true}/M_{\sun})=14.05$.\\
\indent The 943 clusters are separated into two samples according to whether either the DS or Kappa tests detected substructure or not. Of the 943 clusters, dynamical substructure was detected in 255 clusters. PTE values of both the Kappa and DS test for individual clusters can be found in Figure~\ref{fig:PTE_values} and the mass--richness relation of the substructured and non-substructured sample is shown as a red and black solid line respectively in Figure~\ref{fig:SAM2_SW_GCMRP_IV_logM200c_logNgal} in the appendix.\\
\indent The frequency of significant dynamical substructure in our cluster sample is $\sim27\%$. We note that the frequency of significant dynamical substructure varies significantly for observational cluster samples in the literature, with fractions of substructure detected in samples being as low as $\sim15\%$ (e.g., \citealt{1996astro.ph.12206G}), and as high as $\sim80\%$ (e.g., \citealt{2013ApJ...767..102W}). This variation in the fraction of highly substructured clusters is attributed to factors such as differences in the algorithms used to detect substructure and the characteristics of the cluster samples themselves (for example, survey depth, number of galaxies for which there are spectroscopic redshifts available; \citealt{2001MNRAS.320...49K}; \citealt{2004MNRAS.352..605B}; \citealt{2007A&A...470...39R}). In Figure~\ref{fig:subs_vs_mass} in the appendix, we show the prevalence of highly substructured clusters as a function of log true mass, which we find increases for higher mass clusters. This trend is also identified in several observational studies which employ different dynamical substructure tests (e.g.,  \citealt{2017MNRAS.467.3268R}; \citealt{2017AJ....154...96D}).\\
\indent When assessing differences in cluster mass reconstruction of two samples, it is important to control by cluster mass, especially as cluster mass estimation technique performance is often mass dependent. We ensure that the median mass of the two samples are similar by binning the clusters in each sample into seven linearly spaced log true mass bins. We then randomly select the minimum number of distinct clusters in a given mass bin of the two samples. We do this iteratively ($N=200$ iterations), resulting in $N$ sub-samples of substructured clusters and $N$ sub-samples of non-substructured clusters. These subsamples are controlled to have median mass values close to the median mass of the substructured cluster sample (log $(M_{\rm 200c, true}/M_{\sun})=14.13$). As the sample of highly-substructured clusters is smaller, each $N$ sub-samples of substructured clusters typically consists of the same clusters, whereas each $N$ sub-samples of the non-substructured clusters often consists of different clusters within each mass bin.\\
\indent For each set of $N$ sub-samples of dynamically substructured and non-substructured clusters, we quantify differences between the two samples in terms of scaling relations between the true and recovered cluster masses. The first statistic we assess is the scatter in the recovered mass, $\sigma_{M_{\rm Rec}}$, which delivers a measure of the scatter about the fit between true and recovered mass. The second parameter is the slope in the relation between recovered and true underlying mass, $s$, and the third parameter is the amplitude of the fit at the pivot mass, $a$. These statistics are computed by performing a likelihood-fitting analysis on these 400 subsamples, assuming a model where there is a linear relationship between the recovered and true log mass and residual offsets in the recovered mass are drawn from a normal distribution: $\log M_{\rm Rec} = (a+\log M_{\rm Pivot}) + s(\log M_{\rm True}-\log M_{\rm Pivot})+ e$, where $a$, $s$ and $e$ are the amplitude (or normalization), slope and scatter, which includes measurement and model errors in addition to intrinsic scatter (induced by the different physical conditions of each cluster).\\
\indent This analysis is similar to that in \citet{2015MNRAS.449.1897O} and we refer the reader there for more detail. To summarise this approach, we compute a likelihood that is a sum of the probability of obtaining the data point assuming it is drawn from a `good' distribution and the probability of obtaining the data point assuming it is drawn from a `bad' outlier distribution, to try to ensure that the scatter value is not affected by a small number of extreme outliers (see \citealt{2010arXiv1008.4686H} for more details). The components of this likelihood are weighted by the probability that any given point belongs to either of these distributions:
\begin{eqnarray} 
\mathcal{L} &\!\!\!\!=\!\!\!\!& \prod_{i=1,N} p_i \nonumber \\
p_i &\!\!\!\!=\!\!\!\!& \left[(1-P_{\rm b}) P(\log M_{{\rm Rec},i} |  \log
M_{{\rm True},i}
,\sigma_{\log M_{{\rm Rec}, i}},s,a) \right.
\nonumber \\
&\mbox{}& \qquad 
+ \left.
P_{\rm b} P(\log M_{{\rm Rec}, i}|\log M_{{\rm True},i},\sigma_{\rm
  outlier},s,a) \right] \ .
\end{eqnarray}
 $P_{\rm b}$ represents the posterior fraction of objects belonging to the `bad'
outlier distribution, $\sigma_{M_{\rm Rec, i}}$ is the variance of the `good'
distribution and $s$ and $a$ are the slope and amplitude of the fit
respectively. We fix the variance of the `bad' outlier distribution to a very large number with a prior that the variance of the `good' distribution
must always be smaller than the variance of the `bad' distribution. We adopt flat priors for the variance of the `good' distribution, the slope and the amplitude. The probability that $N$ data points belong to a `bad' outlier distribution must be between zero and one. We note that we have performed the analysis with alternative priors (Jeffreys priors), and our results do not change significantly. We utilise Markov Chain Monte Carlo (MCMC) techniques to efficiently sample our parameter space and produce posterior probability distributions for the parameters described above. We use the parallel-tempered MCMC sampler {\sc emcee}  which employs several ensembles of \emph{walkers} at different \emph{temperatures} to explore our parameter space \citep{2013PASP..125..306F}.\\
\indent Employing walkers at different `temperatures' where the
likelihood is modified, enables walkers to easily explore different local maxima, preventing walkers becoming stuck at regions of local instead of global maxima in the case of a multi-modal likelihood. In this analysis, we employ 50 walkers at 5 temperatures and perform 2200 iterations, including a `burn-in' of 1000 iterations that are discarded. In total, $50\times5\times2200 = 5\;500\;000$
points in parameter space are sampled for each method and input catalogue. Figures of the marginalised probability distributions of parameters for all methods are available upon request.\\
\indent We perform the analysis described below for each $N$ sub-sample of highly-substructured and $N$ sub-sample non-substructured clusters and then compute the median of these output parameters of all subsamples.
\section{Results} 
\noindent The goals of this study are assess the extent to which galaxy-based cluster mass estimation techniques are sensitive to the presence of significant dynamical substructure, and ultimately, whether cluster cosmology studies utilising galaxy-based mass estimation should look to exclude dynamically substructured clusters from their samples. We apply observational dynamical substructure tests to our sample of 943 mock clusters to separate our sample into highly-substructured and non-substructured clusters. We then assess whether commonly-used galaxy-based cluster mass estimation techniques perform differently on these two samples. In the following subsections, we discuss the impact of significant dynamical substructure on cluster mass estimation using three key statistics with which we assess how the cluster mass estimation techniques perform. These statistics are the scatter in the relation between recovered and true mass, the amplitude in the relation between recovered and true mass and finally, the mass-dependence i.e., slope in the relation between recovered and true mass.
\subsection{Impact of dynamical substructure on scatter}
\begin{figure}
 \centering
  \includegraphics[trim = 35mm 0mm 0mm 0mm, clip, width=0.66\textwidth]{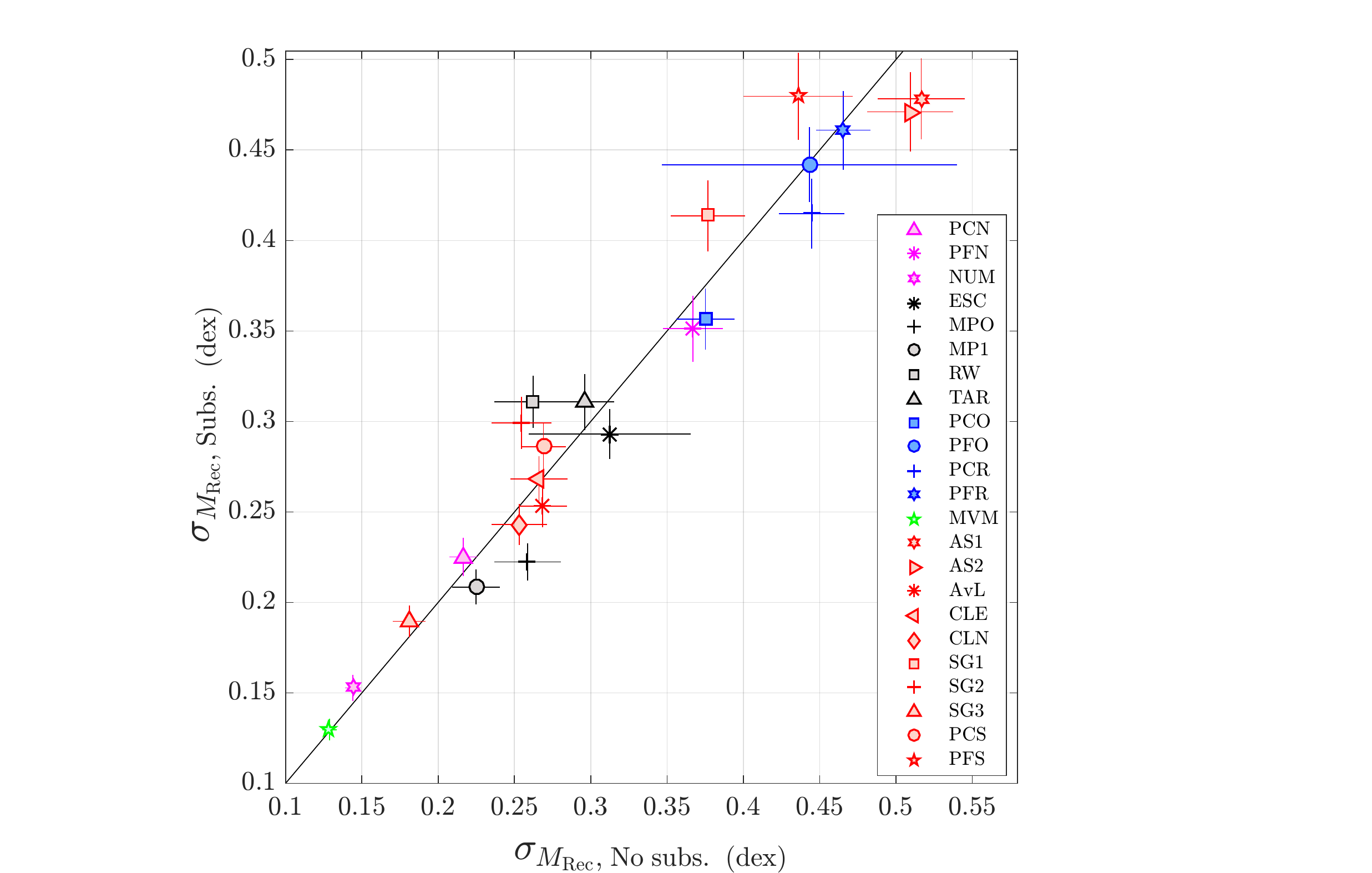}
 \caption{The median scatter in recovered mass produced by each cluster mass estimation technique for the sample of clusters with significant dynamical substructure versus the median scatter in recovered mass for the sample of clusters without significant dynamical substructure. The solid black line represents a 1:1 relation.}
\label{fig:SAM2_SW_median_sigma_rec_DS_OR_Kappa_subs_vs_non_200_v11_paper}
\end{figure}

Figure~\ref{fig:SAM2_SW_median_sigma_rec_DS_OR_Kappa_subs_vs_non_200_v11_paper} depicts the median scatter in recovered mass produced by each cluster mass estimation technique for the highly substructured cluster sample versus the median scatter in recovered mass produced by each cluster mass estimation technique for the non-substructured cluster sample. The solid black line represents a 1:1 relation between these two parameters. The colour scheme reflects the approach implemented by each method to deliver a cluster mass from a chosen galaxy membership: magenta (richness), black (phase-space), blue (radial), green (abundance-matching) and red (velocity dispersion). We find methods that produce lower scatter in recovered mass (situated in the left hand corner of Figure~\ref{fig:SAM2_SW_median_sigma_rec_DS_OR_Kappa_subs_vs_non_200_v11_paper}), show little difference in scatter for both highly-substructured and non-substructured cluster samples. The x-axis error bars show the uncertainty in the scatter parameter for non-substructured clusters, which is calculated by taking the standard deviation of the median scatter parameter values from the set of 200 non-substructured cluster samples. The y-axis error bars show the uncertainty in the scatter parameter for substructured clusters. This uncertainty is calculated by adding in quadrature the uncertainty from the standard deviation of the median scatter parameter values from the set of 200 substructured cluster samples to the uncertainty of the MCMC sampling of the scatter parameter (this former uncertainty is very small as the subsamples typically include the same clusters).\\
\indent While certain methods producing higher scatter in recovered mass may produce higher scatter for highly-substructured clusters (on the order of $\sim15\%$), for example, SG1, PFS, we also see that other methods that utilise similar galaxy-based properties, may produce lower scatter for highly-substructured clusters (on the order of up to $\sim10\%$) for example, AS1, AS2 and PCR. We do not see any consistent behaviour in terms of an increase or decrease in scatter for substructured clusters with mass estimation technique type (i.e, richness, phase-space, radial, abundance matching, velocity dispersion).
\subsection{Impact of dynamical substructure on the amplitude}
\begin{figure}
 \centering
 \includegraphics[trim = 40mm 0mm 0mm 6mm, clip, width=0.62\textwidth]{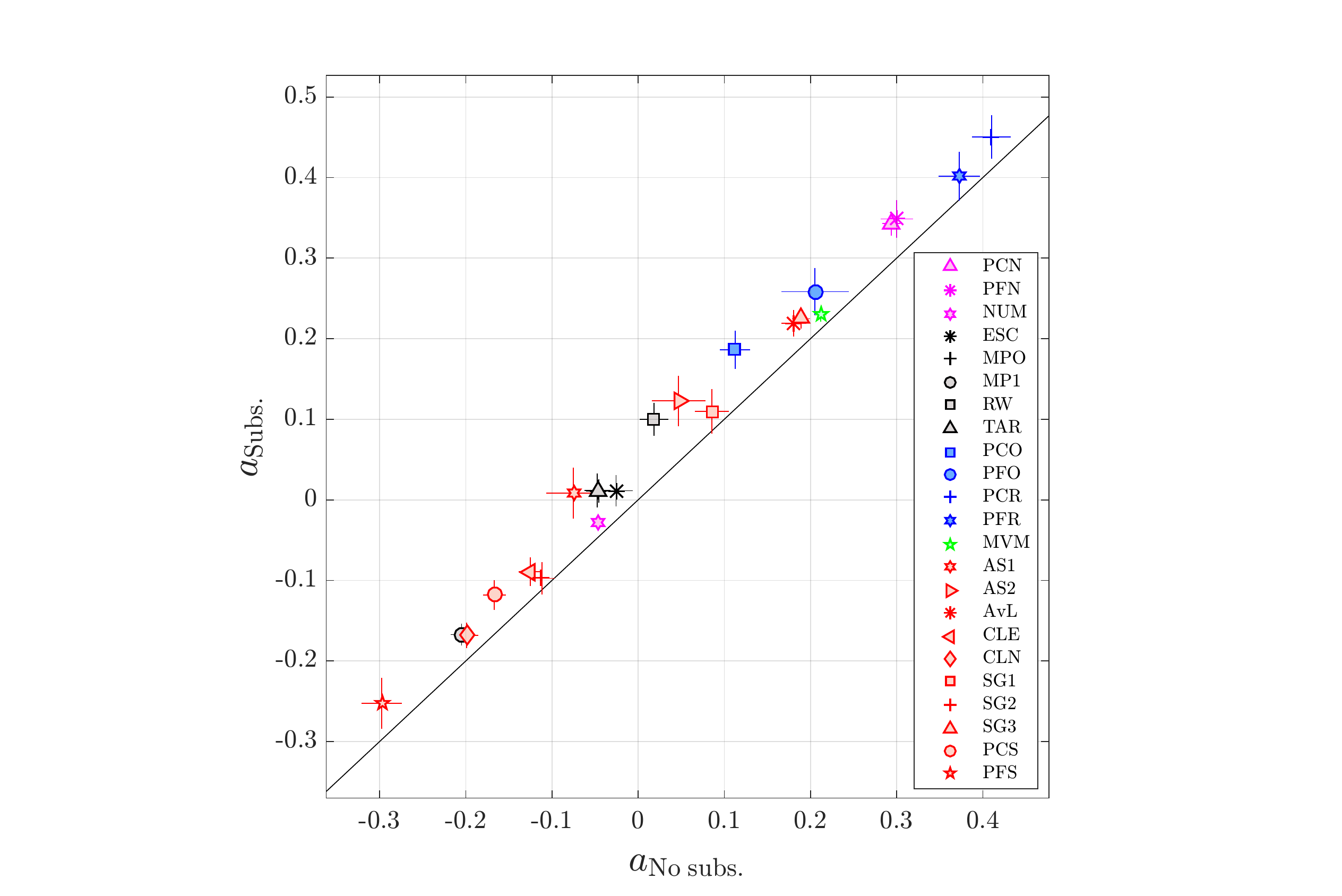}
 \caption{The median amplitude at the pivot mass for the sample of clusters with significant dynamical substructure versus the median amplitude at the pivot mass for the sample of clusters without significant dynamical substructure for each cluster mass estimation technique. The solid black line represents a 1:1 relation. If there were no difference in the amplitudes produced by each method at the pivot mass for the highly-substructured and non-substructured samples, the methods' median amplitude markers would lie on the 1:1 relation.}
\label{fig:SAM2_SW_median_amplitude_DS_OR_Kappa_subs_vs_non_200_v11_paper}
\end{figure}

In addition to scatter, it is important to examine how the presence of significant dynamical substructure affects the amplitude in the relation between recovered and true underlying cluster mass. In this study, we measure the amplitude at the pivot mass which reflects the normalisation of the relation between recovered and true log mass produced by each cluster mass estimation technique. Figure~\ref{fig:SAM2_SW_median_amplitude_DS_OR_Kappa_subs_vs_non_200_v11_paper} shows the median amplitude at the pivot mass of log $M_{\rm 200c, true}=14.13$ for the highly substructured cluster sample versus the median amplitude at the pivot mass produced by each cluster mass estimation technique for the non-substructured cluster sample. The x-axis error bars show the uncertainty in the amplitude parameter for non-substructured clusters, which is calculated by taking the standard deviation of the median amplitude parameter values from the set of 200 non-substructured cluster samples. The y-axis error bars show the uncertainty in the amplitude parameter for substructured clusters. This uncertainty is calculated by adding in quadrature the uncertainty from the standard deviation of the median amplitude parameter values from the set of 200 substructured cluster samples to the uncertainty of the MCMC sampling of the amplitude parameter (this former uncertainty is very small as the subsamples typically include the same clusters).\\
\indent If there were no difference in the biases produced by each method at the pivot mass for the highly-substructured and non-substructured samples, the methods' median amplitude markers would lie on the 1:1 relation. Instead, we see a systematic increase in the amplitude for all techniques for the highly-substructured sample compared to the non-substructured sample. For some methods that underestimate cluster mass in general, for example, velocity dispersion methods PFS, CLN, PCS, CLE and phase-space method MP1, this systematic shift brings the amplitude value slightly closer to zero, and more comparable to the true underlying cluster mass.\\
\indent For the methods that significantly overestimate cluster mass at the pivot mass, for example, radial based methods PCR, PFR, PFO and richness methods PCN and PFN, the amplitude values increase and are brought further away from the average true underlying cluster mass. The median difference for all methods in the amplitude at the pivot mass for the highly substructured cluster sample versus non-substructured cluster samples, $\Delta a = a_{\rm {Subs.}}-a_{\rm{No\;subs.}}$, is $\Delta a = 0.040$ dex ($\sim9.7\%$). We note that this value reflects the average difference in amplitude for all techniques for samples that comprise of only highly-substructured clusters versus only non-substructured clusters.\\
\indent In the likely case that `relaxed', non-substructured clusters are used to calibrate scaling relations with mass, and these scaling relations are then applied to a larger sample of clusters that include both substructured and non-substructured clusters, this bias will likely be smaller. We repeat the MCMC likelihood analysis to compare the amplitude for non-substructured clusters compared to all 943 clusters (substructured and non-substructured clusters) and find a median difference in amplitude of $\Delta a = 0.029$ dex ($\sim6.9\%$) at the pivot mass of log $M_{\rm 200c, true}=14.13$. Note that the median mass of these two samples is kept within $\sim0.009$~dex of each other by subsampling as for the analysis described in Section~4.3.\\
\indent We note that the difference in amplitude increases to $\Delta a =0.067$ dex ($\sim16.8\%$), when we re-run the analysis with a more conservative DS and Kappa test PTE threshold to PTE $\leq 0.01$. This increase in bias likely arises from the increased `purity' in the substructured sample, due to the more pronounced substructure. In addition, we also find that the magnitude of the measured bias increases to $0.06$ dex ($\sim14.6\%$) when we re-run the analysis for the case the mock cluster sample is split into substructured and non-substructured clusters if only $\it both$ the DS and Kappa test classify the cluster as highly substructured (with PTE $\leq 0.05$), as opposed to if $\it either$ the DS and Kappa test classify the cluster as highly substructured.
\subsection{Impact of dynamical substructure on slope}
\begin{figure}
 \centering
 \includegraphics[trim = 9mm 0mm 0mm 0mm, clip, width=0.53\textwidth]{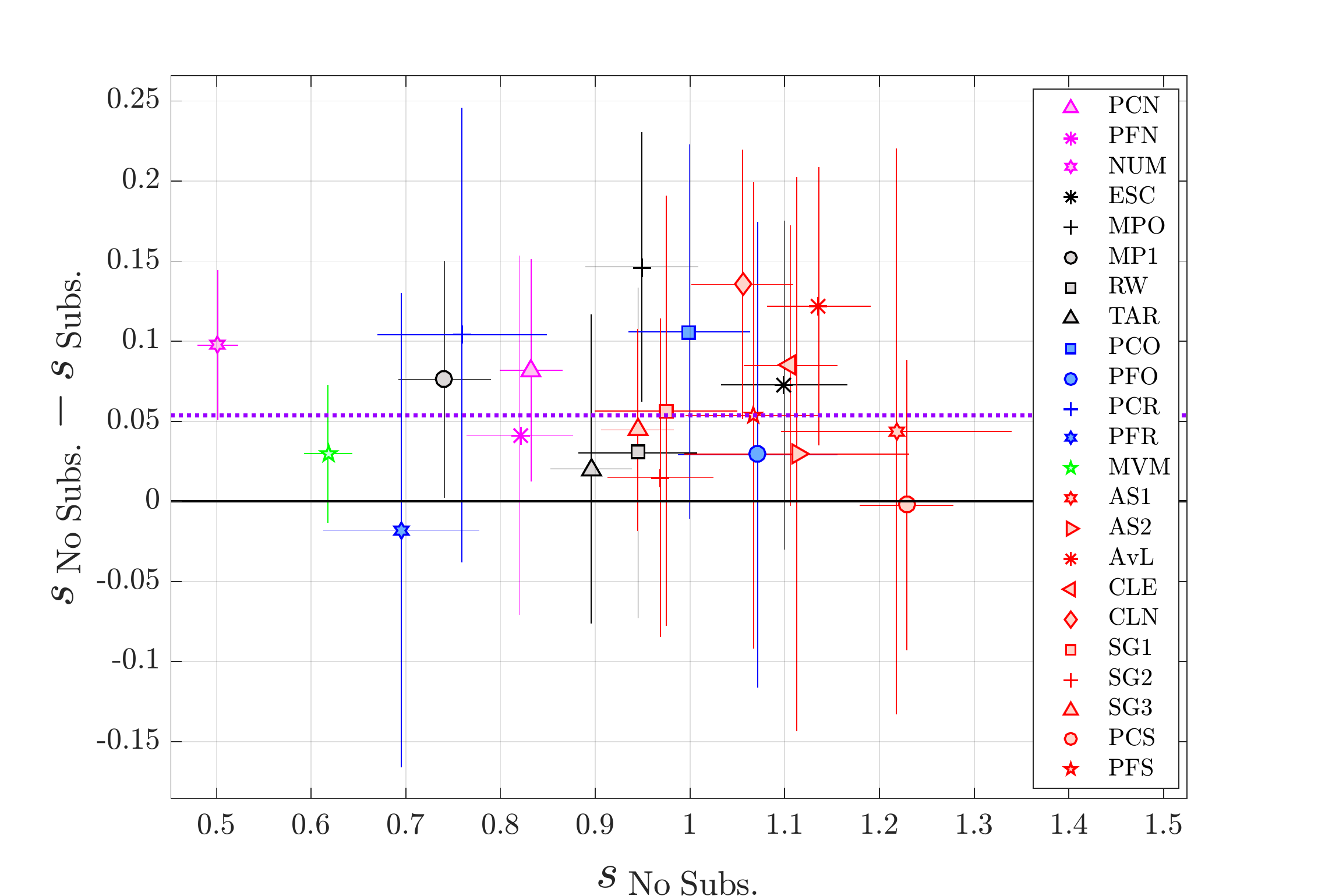}
 \caption{The difference in the slope of the relation between recovered and true log mass produced by each cluster mass estimation technique for the sample of non-substructured clusters to the sample of highly-substructured clusters versus the slope for the sample of non-substructured clusters. The solid black line represents no difference in slope produced by these methods for these two different samples. The dotted red line represents the median difference in the slopes for the two samples for all methods (0.054~dex, $\sim13\%$).  }
\label{fig:SAM2_SW_median_normilised_slope_vs_non_DS_OR_Kappa_subs_slope_v11_paper}
\end{figure}

We now examine the mass dependence in cluster mass reconstruction, to deduce whether methods under- or over-estimate cluster mass differently for lower and higher mass clusters if they have significant dynamical substructure. Figure~\ref{fig:SAM2_SW_median_normilised_slope_vs_non_DS_OR_Kappa_subs_slope_v11_paper} shows the difference in the slope of the relation between recovered and true log mass produced by each cluster mass estimation technique for the sample of non-substructured clusters to the sample of highly-substructured clusters versus the slope for the non-substructured clusters. The solid black line represents no difference in slope produced by these methods for these two different samples. The dotted purple line represents the median difference in the slopes for the two samples for all methods (0.054~dex,$\sim13\%$). The x-axis error bars show the uncertainty in the slope parameter for non-substructured clusters, which is calculated by taking the standard deviation of the median slope parameter values from the set of 200 non-substructured cluster samples. The y-axis error bars show the uncertainty in the difference in slopes, which is calculated by adding in quadrature the uncertainty in the slope for non-substructured clusters and the uncertainty in the slope for substructured clusters. 
The uncertainty in the slope parameter for substructured clusters is calculated by adding in quadrature the uncertainty from the standard deviation of the median slope parameter values from the set of 200 substructured cluster samples to the uncertainty of the MCMC sampling of the slope parameter (this former uncertainty is very small as the subsamples typically include the same clusters).\\ 
\indent We see that the majority of methods produce a slightly flatter slope of the relation between recovered and true log mass for highly-substructured clusters. This behaviour indicates that the masses of higher mass clusters are underestimated and the masses of lower mass clusters are overestimated compared to that of clusters around the pivot mass. Since we also find that cluster masses are systematically biased high at the pivot mass (Section~5.2), these two effects are likely to result in high mass clusters having relatively unbiased masses, while the masses of low mass clusters will likely be biased very high. This is indicated by the linear fit to the substructured clusters in Figure~\ref{fig:SAM2_SW_median_delta_M_Mtrue_clusters} in the appendix, which shows the median difference in recovered and true cluster mass for all 23 mass estimation techniques.
This flattening of the slope also demonstrates that magnitude of the bias in recovered mass ($\sim 10\%$ at the pivot mass) does depend on the underlying cluster mass. For example, if a method systematically overestimated cluster mass by $\sim10\%$ for clusters with a true mass of $\sim$log $M_{\rm 200c, true}=14.13$, that method would likely overestimate the masses of clusters log $M_{\rm 200c, true}<14.13$ to a greater extent.\\
\indent Whilst we see a general trend to flatter slopes between the recovered and true cluster mass, methods that utilise the same galaxy population property to reconstruction, for example the velocity dispersion (red markers), are not all affected in the same manner. This further highlights the diversity in performance of methods which use the same galaxy property as a mass proxy. 
\section{Discussion}
The main objectives of this study are to deduce whether the inclusion of clusters with significant dynamical substructure will produce biases in cluster mass estimation and explore how these biases will impact both galaxy-based cluster cosmology studies and galaxy evolution studies that characterise galaxy environment by cluster mass. Reassuringly, for the majority of galaxy-based techniques with lower intrinsic scatter, we see little difference in the scatter in the recovered versus underlying mass for non-substructured and substructured clusters. However, as shown in Figure~\ref{fig:SAM2_SW_median_amplitude_DS_OR_Kappa_subs_vs_non_200_v11_paper} and 
Figure~\ref{fig:SAM2_SW_median_normilised_slope_vs_non_DS_OR_Kappa_subs_slope_v11_paper}, the presence of significant dynamical substructure does indeed bias the amplitude and the slope in the relation between true underlying mass and estimated mass for all 23 cluster mass estimation techniques in this study.\\
\indent The direction of this bias, i.e., the increase in estimated cluster mass compared to the true underlying mass for highly dynamically substructured clusters, is qualitatively in agreement with both \citet{1990A&A...237..319P}; \citet{1996ApJS..104....1P} and \citet{2006A&A...456...23B}, who find that in the case of virial-based cluster mass specifically, masses are overestimated for N-body simulations of merging clusters. For a more direct comparison, we apply our analysis to the simulated data-set of 62 cluster-sized haloes in 3 projections from \citet{2006A&A...456...23B}. For clusters that are highly substructured in projected phase-space compared to unsubstructured, we measure a bias between the recovered virial-based mass to true mass of (0.12~dex, $\sim32\%$) at a pivot mass of log $M_{\rm 200c, true}=14.13$, which is consistent with the bias we see for several methods. In addition, we perform a both a two-sample KS test and a two-sample Anderson-Darling test on this data-set which rejects the null hypothesis that the recovered virial-based masses of substructured and non-substructured clusters are drawn from the same underlying continuous distribution (with PTE's of 0.0029 and 0.0038 respectively). \\
\indent The analyses described above indicate a bias in virial-based cluster mass estimation. We highlight that the bias we find is prevalent in all 23 galaxy-based techniques which encompass richness, projected phase-space, radial and abundance matching-based techniques. For richness-based techniques, this bias could be partially explained by differences in the stacked mass--richness relation for the substructured and non-substructured samples. A linear fit to the stacked samples, for example, delivers an increase in log mass of 0.07~dex at fixed $N_{\rm gal}$ of 40. However, we see substructures causing a consistent bias across all galaxy-based techniques that do not reconstruct mass from galaxy number counts.\\
\indent The exact impact of this substructure-induced mass bias will be highly dependent on the underlying properties of individual cluster samples; however, we wish to qualitatively deduce the relevance of this bias. The most direct channel of propagating the bias into the estimates of cosmological parameters occurs when a cluster sample used for calibrating a mass scaling relation includes galaxy clusters with a different degree of substructure than the entire cluster sample used for cosmological inference. Considering the most extreme case, the calibration sample may consist of fully relaxed, non-substructured clusters. The primary effect of this observational strategy would be a shift of the observed mass function along the mass axis which in turn would cause a biased measurement of $\Omega_{m}$ and $\sigma_{8}$. \begin{figure}
 \centering
 \includegraphics[trim = 25mm 0mm 0mm 0mm, clip, width=0.55\textwidth]{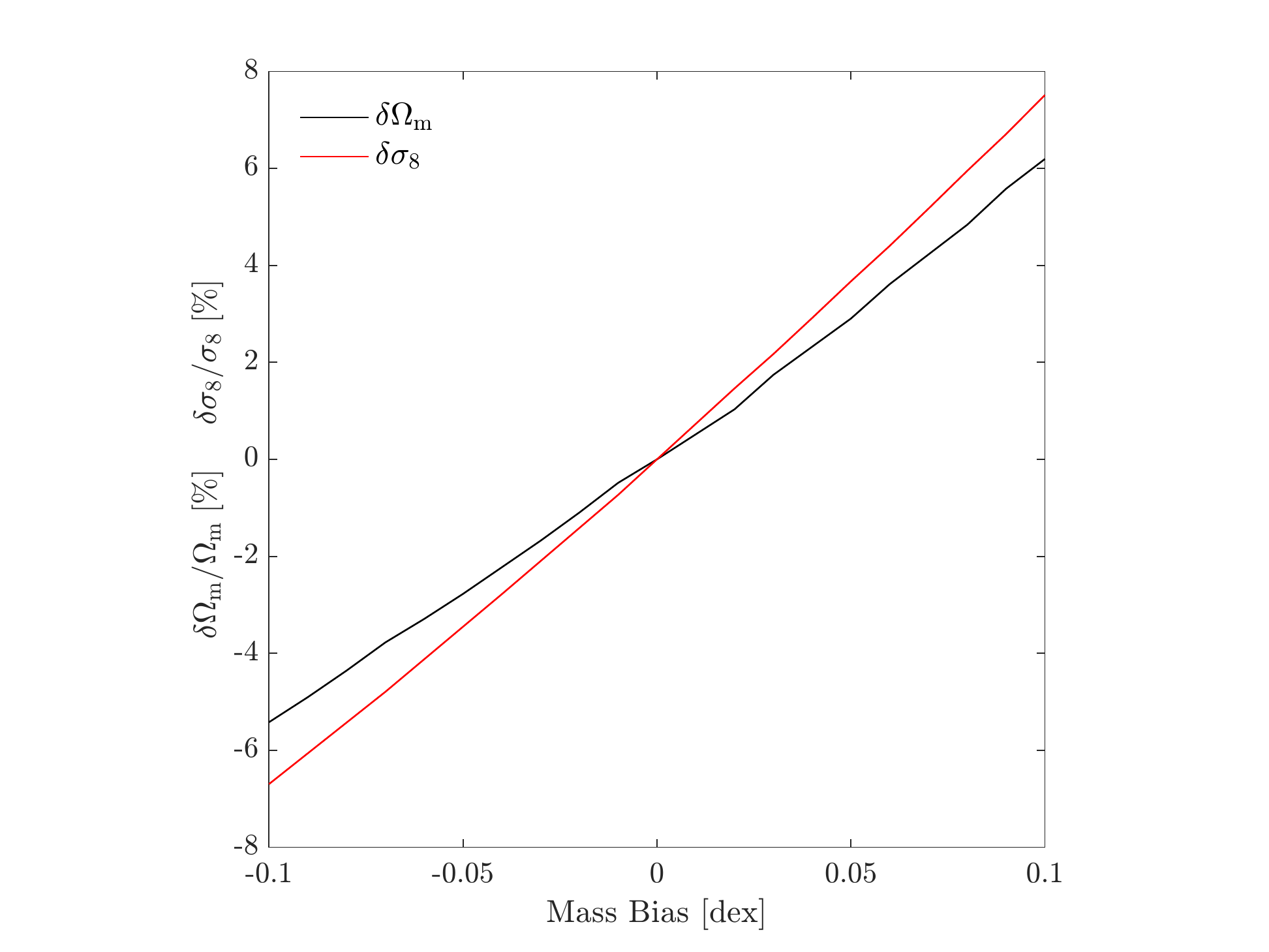}
 \caption{The percentage difference in $\Omega_{m}$ and $\sigma_{8}$ found when fitting a $\Lambda$CDM mass function with Planck parameters when shifting the mass function in log $M_{\rm 200c}$ by a range of values between $-0.1$ and $0.1$ dex.}
\label{fig:bias-Om-sigma8}
\end{figure}A simple way to estimate the potential relative bias in the two cosmological parameters is to determine the two cosmological parameters for which the corresponding mass function matches the mass function computed for a fixed, fiducial cosmology, but shifted along the mass axis by a range of mass biases. In our calculation we adopt a Planck cosmology (\citealt{2016A&A...594A..13P}) with $\Omega_{m}=0.31$ and $\sigma_{8}=0.83$ as a reference model and a universal fitting formula for the mass function from \citet{2008ApJ...688..709T}. Figure~\ref{fig:bias-Om-sigma8} shows the results for a range of mass biases. Interestingly, the error on $\Omega_{m}$ and $\sigma_{8}$ is on the same order as the error on the current leading constraints from CMB-based cosmology studies such as \citet{2016A&A...594A..13P} and is slightly lower than the error produced by weak lensing cluster cosmology studies such as \citet{2015MNRAS.446.2205M} and SZ-based cluster cosmology studies \citep{2016ApJ...832...95D}. We note that this is for the extreme case that the calibration sample is non-substructured, and the majority of the full sample of clusters are highly substructured. In the more realistic case that the contamination of highly substructured clusters in a given survey is typical to the fraction we observe in our simulated sample, $\sim 27\%$, the systematic error on is likely on $\Omega_{m}$ and $\sigma_{8}$ is on the order of $\sim1\%$.

\section{Conclusions}
\label{sec: Conclusions}
In this paper, we examine whether the masses of dynamically disturbed clusters can be measured to the same accuracy and precision as dynamically relaxed clusters with a variety of commonly-used galaxy-based cluster mass estimation techniques. We aim to understand whether scaling relations between multi-wavelength mass estimation techniques would differ for highly substructured and non-substructured clusters, and to that end, whether dynamically young clusters should be excluded from future galaxy-based cluster cosmology samples.  The main results are as follows:
\mbox{}
\begin{enumerate}
\item  For the majority of galaxy-based techniques with lower intrinsic scatter, we see little difference in the scatter in the recovered versus underlying mass for non-substructured and substructured clusters.
\item We see a systematic increase in the measured amplitude at the median mass of the sample for all techniques for the highly-substructured sample compared to the non-substructured sample. This means that for the same given underlying true cluster mass, all cluster mass measurement techniques will, on average, overestimate the mass of a cluster if it has significant dynamical substructure compared to a dynamically relaxed cluster. This systematic bias for all cluster mass estimation techniques is, on average, $\sim10\%$ for clusters around log $M_{\rm 200c}=14.13$. It should be noted that for some methods which underestimate cluster mass in general, this systematic increase in amplitude brings measured cluster masses closer to the true underlying cluster mass, and vice versa.
\item We find that the bias in cluster mass for dynamically disturbed clusters is indeed mass dependent. Typically, the slope of the relation between recovered and true cluster mass is flatter for the sample of highly substructured clusters. A flatter slope indicates that the masses of higher mass clusters are underestimated and the masses of lower mass clusters are overestimated in comparison to the reconstructed masses of clusters at the median mass of the sample ($\sim$log $M_{\rm 200c}=14.13$). The combination of a flatter slope and a positive bias in amplitude at the pivot mass indicate that the reconstructed masses of clusters at the high mass end are likely to be only minimally biased, whereas the reconstructed masses of clusters at the low mass end are biased even higher (for group-sized systems, this bias is $\ga20\%$ for $\la10^{13.5}$).
\item For the purpose of improving accurate deductions of cosmological parameters from future galaxy-based cluster cosmology samples, or accurate characterisation of environment for galaxy evolution studies, we recommend the dynamical state of a cluster sample is classified to identify whether masses of the dynamically substructured clusters will be systematically overestimated. In the case of using cluster mass scaling relations to estimate masses of another cluster sample, we advise that the underlying dynamical characteristics of the cluster sample used to calibrate the scaling relation is similar to that of the cluster sample the scaling relation is applied to.\end{enumerate}
\section*{Acknowledgments}
The authors would like to thank numerous people for useful discussions,
including Matt Owers, Rene\'{e} Hlo\u{z}ek, Irene Pintos-Castro and Joanne Cohn. We would like to acknowledge funding from the Science and Technology Facilities Council (STFC). DC would like to thank the
Australian Research Council for receipt of a QEII Research
Fellowship. The authors would like to express special thanks to the Instituto de Fisica Teorica (IFT-UAM/CSIC in Madrid) for its hospitality and support, via the Centro de Excelencia Severo Ochoa Program under Grant No. SEV-2012-0249, during the three week workshop ``nIFTy Cosmology" where this work developed. We further acknowledge the financial support of the University of Western 2014 Australia Research Collaboration Award for ``Fast Approximate Synthetic Universes for the SKA", the ARC Centre of Excellence for All Sky Astrophysics (CAASTRO) grant number CE110001020, and the two ARC Discovery Projects DP130100117 and DP140100198. We also recognise support from the Universidad Autonoma de Madrid (UAM) for the workshop infrastructure. RAS acknowledges support from the NSF grant AST-1055081. CS acknowledges support from the European Research Council under FP7 grant number 279396. 
\bibliographystyle{mn2e}
\bibliography{GCMRP_dynamical_substructure_references}
\section*{Affiliations}
$^{1}$Department of Astronomy $\&$ Astrophysics, University of Toronto, Toronto, Canada\\ 
$^{2}$Kavli Institute
for Particle Astrophysics and Cosmology, Stanford University, 452
Lomita Mall, Stanford, CA 94305-4085, USA\\ 
$^{3}$SLAC National Accelerator Laboratory, Menlo Park, CA 94025, USA\\
$^{4}$Dark Cosmology Centre, Niels Bohr Institute, University of Copenhagen, Juliane Maries Vej 30, DK-2100 Copenhagen, Denmark\\
$^{5}$School of Physics and Astronomy, University of Nottingham, Nottingham, NG7 2RD, UK\\ 
$^{6}$Institut
d\'Astrophysique de Paris (UMR 7095 CNRS $\&$ UPMC), 98 bis Bd Arago, F-75014 Paris, France\\
$^{7}$Department of Astrophysical Sciences, Peyton Hall, Princeton University, Princeton, NJ 08544, USA\\
$^{8}$Leiden Observatory, Leiden University, PO Box 9513, NL-2300 RA Leiden, Netherlands\\
$^{9}$Leibniz-Institut f\"ur Astrophysik Potsdam (AIP), An der Sternwarte 16, 14482 Potsdam, Germany
$^{10}$Tartu Observatory, Observatooriumi 1, 61602 T\~oravere,
Estonia\\ 
$^{11}$INAF-Osservatorio Astronomico di Trieste, via G. B. Tiepolo 11,
34143, Trieste, Italy\\
$^{12}$Instituto Nacional de Pesquisas Espaciais, MCT, S.J. Campos, Brazil\\
$^{13}$University of California, Santa Cruz, Science Communication Program, 1156 High Street, Santa Cruz, CA 95064\\ 
$^{14}$Freelance science journalist, San Diego, CA, USA\\ 
$^{15}$Centre for Astrophysics \& Supercomputing, Swinburne University of
Technology, PO Box 218, Hawthorn, VIC 3122, Australia\\ 
$^{16}$ICRAR, University of Western Australia, 35 Stirling Highway, Crawley, Western Australia 6009, Australia\\
$^{17}$ARC Centre of Excellence for All-Sky Astrophysics (CAASTRO)\\
$^{18}$Department of Physics and Astronomy, Stony Brook University, Stony Brook, NY 11794, USA\\
$^{19}$Faculty of Physics, Ludwig-Maximilians-Universität, Scheinerstr. 1,
81679 Munich, Germany\\
\appendix
\newpage
\begin{table*}

\appendix
\begin{flushleft}
\section{Properties of the Mass Reconstruction Methods}
\end{flushleft}
 \centering
 \caption{Illustration of the member galaxy selection process for all methods. The colour of the acronym for each method colour corresponds to the main galaxy population property used to perform mass estimation richness (magenta), projected phase-space (black), radii (blue), velocity dispersion (red), or abundance matching (green). The second column details how each method selects an initial member galaxy sample, while the third column outlines the member galaxy sample refining process. Finally, the fourth column describes how methods treat interloping galaxies that are not associated with the clusters.}
 \begin{tabular}{p{1.1cm} p{5cm} p{5cm} p{5cm}}
 \toprule
 \multirow{2}{1cm}{\textbf{Methods}}&\multicolumn{3}{c}{Member galaxy selection methodology} \\[1.5ex]
 \cline{2-4}\\[-1.3ex]
 &Initial Galaxy Selection&Refine Membership&Treatment of Interlopers \\[0.15ex]
 \midrule
 \textcolor{magenta}{\textbf{PCN}}&Within $\rm 5\,Mpc$, $\rm 1000\,km\,s^{-1}$&Clipping of $\pm3\,\sigma$, using galaxies within $\rm 1\,Mpc$&Use galaxies at $\rm 3-5 \,Mpc$ to find interloper population to remove \\
 \textcolor{magenta}{\textbf{PFN}}&FOF&No&No \\
 
 \textcolor{magenta}{\textbf{NUM}}&Within $\rm 3\,Mpc$, $\rm 4000\,km\,s^{-1}$
&1) Estimate $R_{\rm 200c}$  from the relationship between $R_{\rm 200c}$  and
 richness deduced from CLE; 2) Select galaxies within $R_{\rm 200c}$  and with
  $|v|<2.7\,\sigma_{\rm los}^{\rm NFW}(R)$&Same as CLE \\
  \textcolor{magenta}{\textbf{RM1}}&Red Sequence&Red Sequence& Probabilistic \\
  \textcolor{magenta}{\textbf{RM2}}&Red Sequence&Red Sequence& Probabilistic \\
\
 \textcolor{black}{\textbf{ESC}}&Within preliminary $R_{\rm 200c}$  estimate and $\rm \pm3500 \,km\,s^{-1}$&Gapper technique&Removed by Gapper technique\\
 
 \textcolor{black}{\textbf{MPO}}&Input from CLN&1) Calculate $R_{\rm 200c}$, $R_{\rm \rho}$, $R_{\rm red}$, $R_{\rm blue}$ by MAMPOSSt method; 2) Select members within radius according to colour&No \\
 
 \textcolor{black}{\textbf{MP1}}&Input from CLN&Same as MPO except colour blind&No \\
 
 \textcolor{black}{\textbf{RW}}&Within $\rm 3\,\,Mpc$, $\rm 4000\,\,km\,s^{-1}$&Within $R_{\rm 200c}$ , $|2\Phi(R)|^{1/2}$, where $R_{\rm 200c}$  obtained iteratively
&No \\

 \textcolor{black}{\textbf{TAR}}&FOF&No&No \\
 
 \textcolor{blue}{\textbf{PCO}}&Input from PCN& Input from PCN&Include interloper contamination in density fitting \\
 
 \textcolor{blue}{\textbf{PFO}}&Input from PFN&Input from PFN&No \\
 
 \textcolor{blue}{\textbf{PCR}}&Input from PCN&Input from PCN&Same as PCN \\
 
 \textcolor{blue}{\textbf{PFR}}&Input from PFN&Input from PFN&No \\

 \textcolor{green}{\textbf{MVM}}&FOF (ellipsoidal search range, centre
of most luminous galaxy)&Increasing mass limits, then FOF, loops until
closure condition&No \\
 
 \textcolor{red}{\textbf{AS1}}&Within $\rm 1\,Mpc$, $\rm 4000\,km\,s^{-1}$, constrained by colour-magnitude relation&Clipping of $\pm3\,\sigma$&Removed by clipping of $\pm3\,\sigma$ \\
 
 \textcolor{red}{\textbf{AS2}}&Within $\rm 1\,Mpc$, $4\rm 000\,km\,s^{-1}$, constrained by colour-magnitude relation&Clipping of $\pm3\,\sigma$&Removed by clipping of $\pm3\,\sigma$ \\
 
 \textcolor{red}{\textbf{AvL}}&Within $2.5\,\sigma_{v}$ and $0.8\,R_{\rm 200}$&Obtain $R_{\rm 200c}$  and $\,\sigma_{v}$ by $\,\sigma$-clipping&Implicit with $\sigma$-clipping \\
 
 \textcolor{red}{\textbf{CLE}}&Within $\rm 3\,Mpc$, $\rm 4000\,km\,s^{-1}$&1)
 Estimate $R_{\rm 200c}$  from the aperture velocity dispersion; 2) Select galaxies within
 $R_{\rm 200c}$  and with $|v|<2.7\,\sigma_{\rm los}^{\rm NFW}(R)$; 3) Iterate steps 1 and 2 until convergence&Obvious interlopers are removed by velocity gap technique, then further treated in iteration by $\sigma$ clipping \\
 
 \textcolor{red}{\textbf{CLN}}&Input from NUM&Same as CLE&Same as CLE \\
 
 \textcolor{red}{\textbf{SG1}}&Within $\rm 4000 \,km\,s^{-1}$&1) Measure $\,\sigma_{\rm gal}$, estimate $M_{\rm 200c}$ and $R_{\rm 200c}$ ; 2) Select galaxies within $R_{\rm 200c}$ ; 3) Iterate steps 1 and 2 until convergence&Shifting gapper with minimum bin size of $\rm 250 \,kpc$ and 15 galaxies; velocity limit $\rm 1000\,km\,s^{-1}$ from main body \\
 
 \textcolor{red}{\textbf{SG2}}&Within $\rm 4000 \,km\,s^{-1}$&1) Measure $\sigma_{\rm gal}$, estimate $M_{\rm 200c}$ and $R_{\rm 200c}$ ; 2) Select galaxies within $R_{\rm 200c}$ ; 3) Iterate steps 1 and 2 until convergence&Shifting gapper with minimum bin size of $\rm 150 \,kpc$ and 10 galaxies; velocity limit $\rm 500 \,km\,s^{-1}$ from main body \\
  \textcolor{red}{\textbf{SG3}}&Within $\rm 2.5\,h^{-1}\,Mpc$ and $\rm   4000\,km\,s^{-1}$. Velocity distribution symmeterised& Measure $\,\sigma_{\rm gal}$, correct for velocity errors, then estimate $M_{\rm 200c}$ and $R_{\rm 200c}$  and apply the
surface pressure term correction&Shifting gapper with minimum bin size of
$\rm 420\,h^{-1} kpc$ and 15 galaxies\\
 \textcolor{red}{\textbf{PCS}}&Input from PCN&Input from PCN&Same as PCN \\
 \textcolor{red}{\textbf{PFS}}&Input from PFN&Input from PFN&No \\
 \bottomrule
 \end{tabular}
 \label{table:appendix_table_1}
\end{table*}
\begin{table*}
\renewcommand\thetable{A2} 
 \centering
 \centering
 \caption{Characteristics of the mass reconstruction process for the methods used in this comparison. The second to sixth columns illustrate whether a method calculates/utilises the velocities, velocity dispersion, radial distance of galaxies from cluster centre, the richness and the projected phase-space information of galaxies respectively. If a method assumed a mass or number density profile it is indicated in columns seven and eight.}
 \begin{tabular}{c c c c c c c c}
 \toprule
 \multirow{2}[4]{*}{\textbf{Methods}}&\multicolumn{7}{c}{Galaxy properties used to obtain group/cluster membership and estimate mass} \\[1.0ex]
 \cline{2-8}
 &Velocities&Velocity dispersion&Radial distance&Richness&Projected phase-space&Mass density profile&Number density profile\\
 \hline
 \textcolor{magenta}{\textbf{PCN}}&Yes&No&No&Yes&No&No&No \\
 \textcolor{magenta}{\textbf{PFN}}&Yes&No&No&Yes&No&No&No\\
 \textcolor{magenta}{\textbf{NUM}}&No&No&No&Yes&Yes&No&No\\
 \textcolor{magenta}{\textbf{RM1}}&No& No& Yes& Yes &No&No& NFW \\
 \textcolor{magenta}{\textbf{RM2}}&No&No& Yes&Yes&No&No& NFW \\
  
 \textcolor{black}{\textbf{ESC}}&Yes&Yes&Yes&No&No&Caustics&No \\
 \textcolor{black}{\textbf{MPO}}&Yes&No&Yes&No&Yes&NFW&NFW\\
 \textcolor{black}{\textbf{MP1}}&Yes&No&Yes&No&Yes&NFW&NFW\\
 \textcolor{black}{\textbf{RW}}&Yes&No&Yes&No&Yes&NFW&NFW\\
 \textcolor{black}{\textbf{TAR}}&Yes&Yes&Yes&No&No& NFW&No\\
 
 \textcolor{blue}{\textbf{PCO}}&Yes&No&No&No&No&NFW&NFW\\
 \textcolor{blue}{\textbf{PFO}}&Yes&No&No&No&No&NFW&NFW\\
 \textcolor{blue}{\textbf{PCR}}&Yes&No&Yes&No&No&No&No\\
 \textcolor{blue}{\textbf{PFR}}&Yes&No&Yes&No&No&No&No \\
  
 \textcolor{green}{\textbf{MVM}}&Yes&Yes&Yes&No&No&NFW&No\\
 
 \textcolor{red}{\textbf{AS1}}&Yes&Yes&No&No&No&No&No\\
 \textcolor{red}{\textbf{AS2}}&Yes&No&Yes&No&Yes&No&No \\
 
 \textcolor{red}{\textbf{AvL}}&Yes&Yes&Yes&No&No&No&No\\
 
 \textcolor{red}{\textbf{CLE}}&Yes&Yes&No&No&No&NFW&NFW\\
 
 \textcolor{red}{\textbf{CLN}}&Yes&Yes&No&No&No&NFW&NFW\\
 
 \textcolor{red}{\textbf{SG1}}&Yes&Yes&Yes&No&No&No&No\\
 
 \textcolor{red}{\textbf{SG2}}&Yes&Yes&Yes&No&No&No&No\\
 
 \textcolor{red}{\textbf{SG3}}&Yes&Yes&Yes&No&No&No&No\\

 \textcolor{red}{\textbf{PCS}}&Yes&Yes&No&No&No&No&No \\
 
 \textcolor{red}{\textbf{PFS}}&Yes&Yes&No&No&No&No&No \\
 \bottomrule \\
 \end{tabular} 
  \label{table:appendix_table_2}
\end{table*}
\vspace{5mm}

\clearpage

\setcounter{figure}{0} \renewcommand{\thefigure}{B\arabic{figure}}
\section{DS and Kappa test PTE values for all clusters}
\begin{figure}
\begin{flushleft}
\end{flushleft}
 \centering
 \includegraphics[trim = 28mm 0mm 0mm 10mm, clip, width=0.56\textwidth]{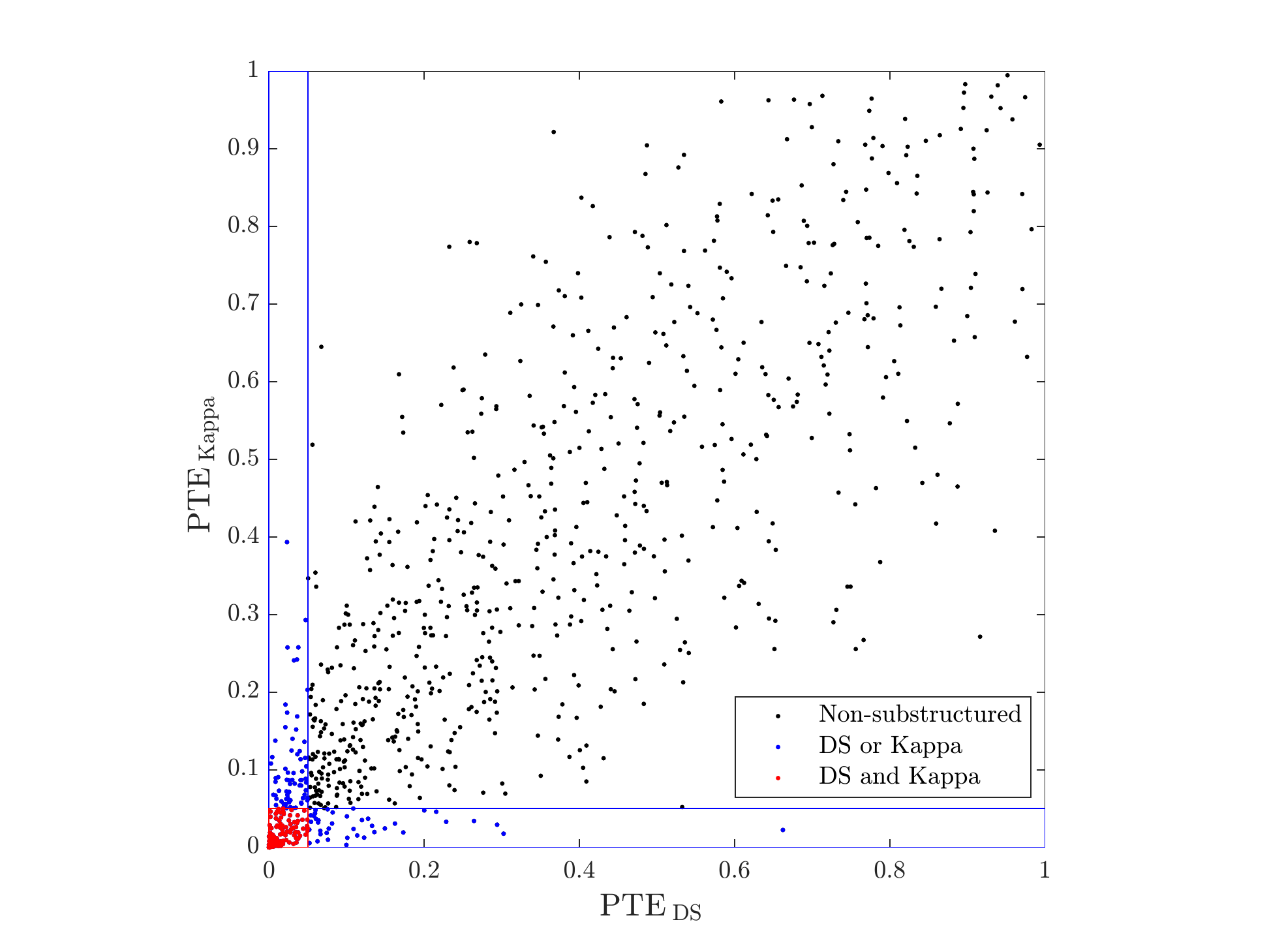}
 \caption{The DS and Kappa test PTE values for the cluster sample. Black symbols indicate clusters that are not defined as highly substructured by either the DS or Kappa test (688 clusters, 73$\%$ of the sample). Blue symbols indicate clusters where either the DS and Kappa test have defined as highly substructured (255 clusters, 27$\%$). The red symbols indicate clusters that have been defined as highly dynamically substructured by both the DS and Kappa test (147 clusters, 15.5$\%$). We note that the DS test detects significant dynamical substructure in 215 clusters, 23$\%$ of the sample. This is a high detection rate than the Kappa test, which finds 187, 20$\%$ of the sample, to be dynamically substructured.}
\label{fig:PTE_values}
\end{figure}
\begin{figure}
\setcounter{figure}{0} \renewcommand{\thefigure}{C\arabic{figure}}

\begin{flushleft}
\section{Dynamical substructure test detection $\&$ cluster mass}
\end{flushleft}
\vspace{1.2mm}

 \centering
 \includegraphics[trim = 28mm 0mm 0mm 0mm, clip, width=0.56\textwidth]{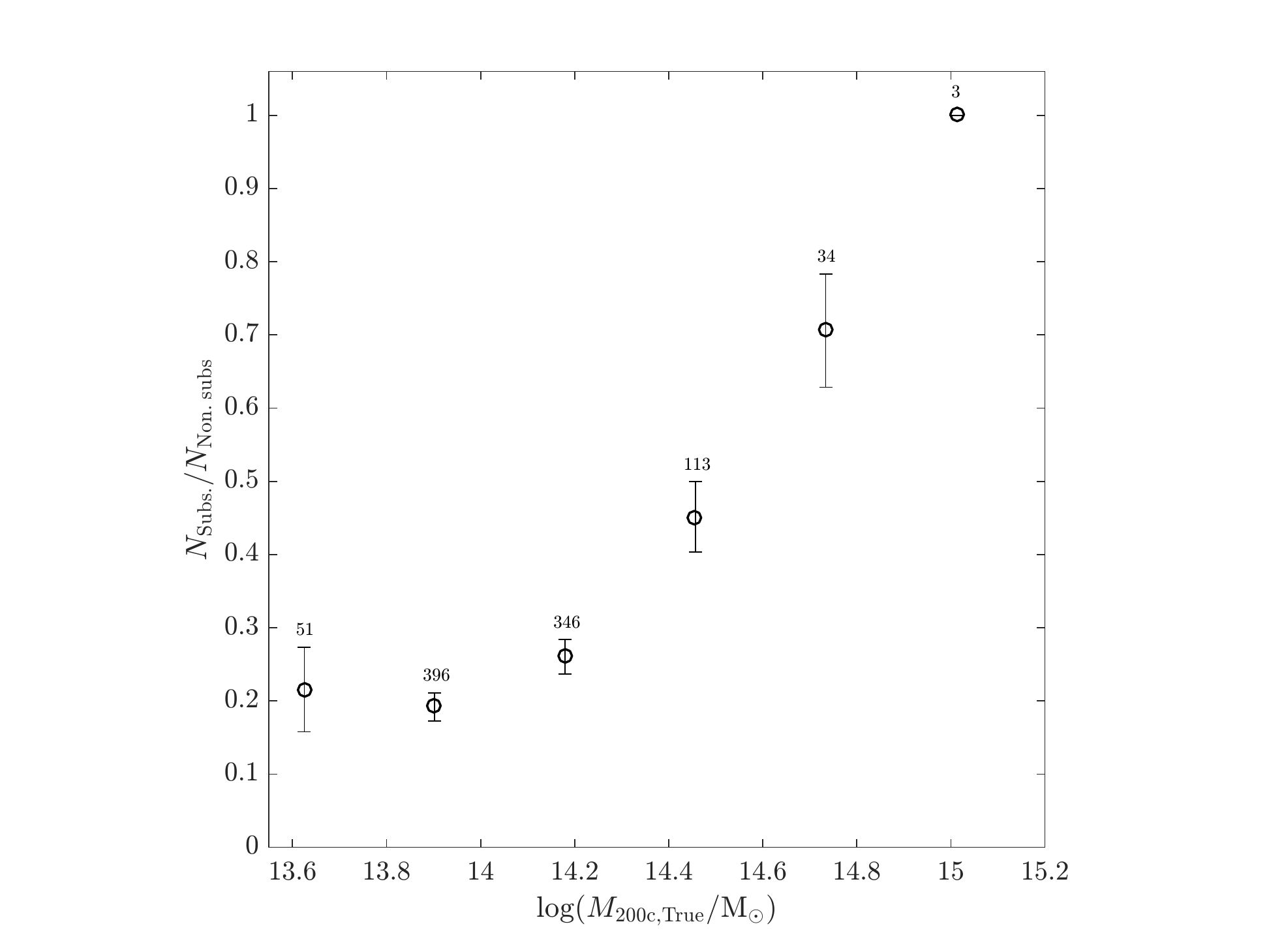}
 \caption{The fraction of highly substructured clusters as a function of log true mass, where clusters are deemed substructured if either the DS or Kappa test detects significant dynamical substructure. The clusters are binned into seven linearly spaced log true mass bins. The error bars represent the standard deviation of a set of fractions calculated by randomly sampling the data with replacement ($n=500$ iterations). The DS and Kappa test detects higher fractions of clusters with substructure as a function of cluster mass (and hence richness). This trend of dynamically disturbed clusters having higher masses is also identified in several observational studies which use different dynamical substructure tests (e.g.,  \citealt{2017MNRAS.467.3268R}; \citealt{2017AJ....154...96D}).}
\label{fig:subs_vs_mass}
\end{figure}
\setcounter{figure}{0} \renewcommand{\thefigure}{D\arabic{figure}}

\begin{figure}
\begin{flushleft}
\section{The richness -- mass relation of the SAM2 mock cluster catalogue}
\end{flushleft}
 \centering
 \includegraphics[trim = 28mm 0mm 0mm 0mm, clip, width=0.56\textwidth]{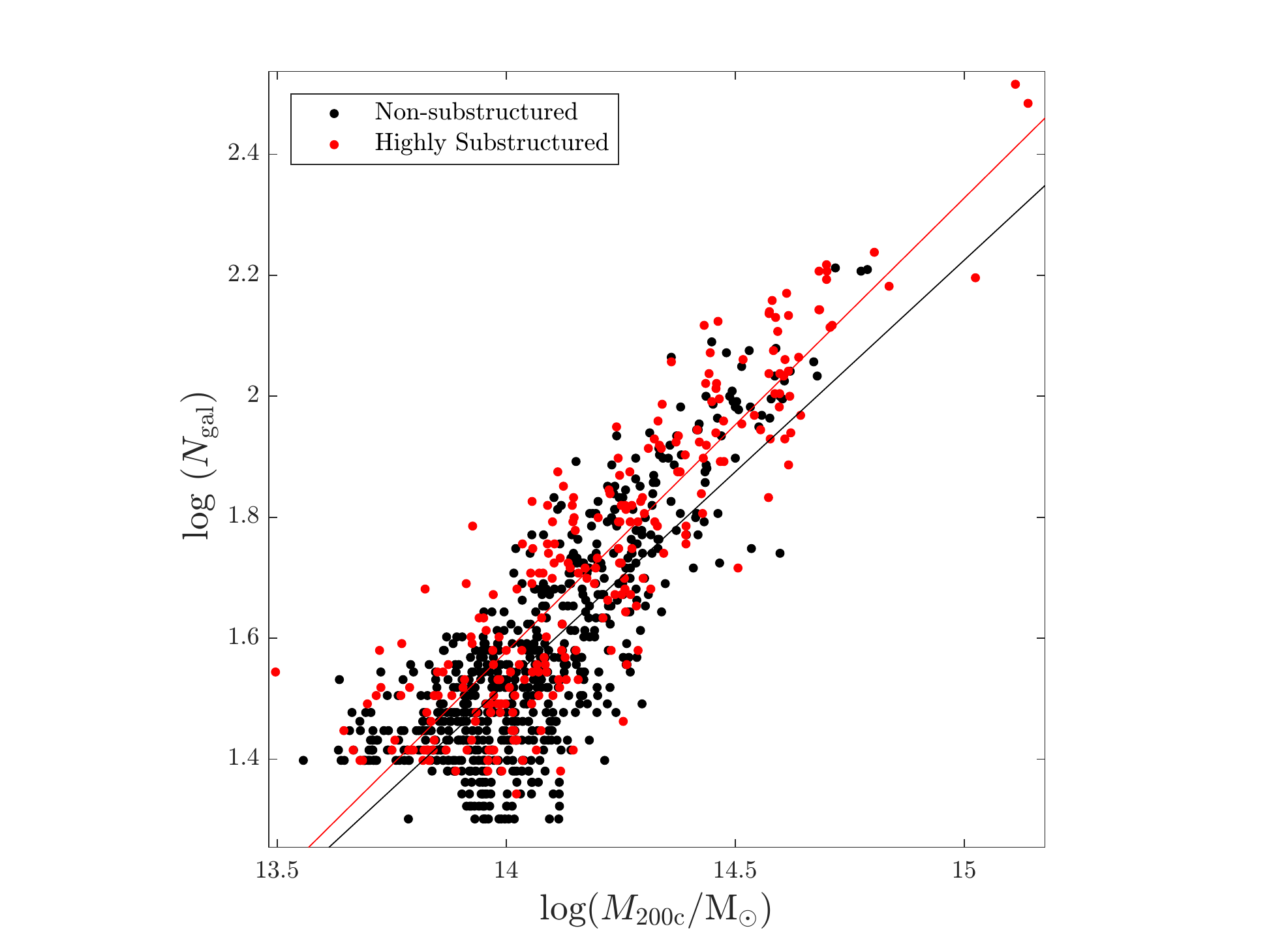}
 \caption{The richness versus mass of the 943 groups/clusters of the input SAM2 catalogues. Clusters that deemed as highly dynamically substructured by the DS or Kappa test are denoted as red circles, and the non-substructured clusters denoted are by black circles. The red line reflects a linear fit (described in Section~4.3) to the richness--mass relation for the substructured clusters of $\rm{log}(N_{\rm gal}) = 0.75\;(\rm{log}(M_{\rm 200c})-14.126)+1.67$. The black line reflects a linear fit to the richness--mass relation for the non-substructured clusters of $\rm{log}(N_{\rm gal}) = 0.70\;(\rm{log}(M_{\rm 200c})-14.126)+1.61$. The intrinsic scatter of the richness versus mass relation of all 943 SAM2 clusters is 0.12 dex. We note that the linear fit parameters are also very similar to those deduced by performing simple linear fit.}
\label{fig:SAM2_SW_GCMRP_IV_logM200c_logNgal}
\end{figure}

\begin{figure}
\setcounter{figure}{0} \renewcommand{\thefigure}{E\arabic{figure}}
\begin{flushleft}
\section{The median difference in recovered mass for all 23 techniques}
\end{flushleft}
 \centering
 \includegraphics[trim = 28mm 0mm 0mm 0mm, clip, width=0.56\textwidth]{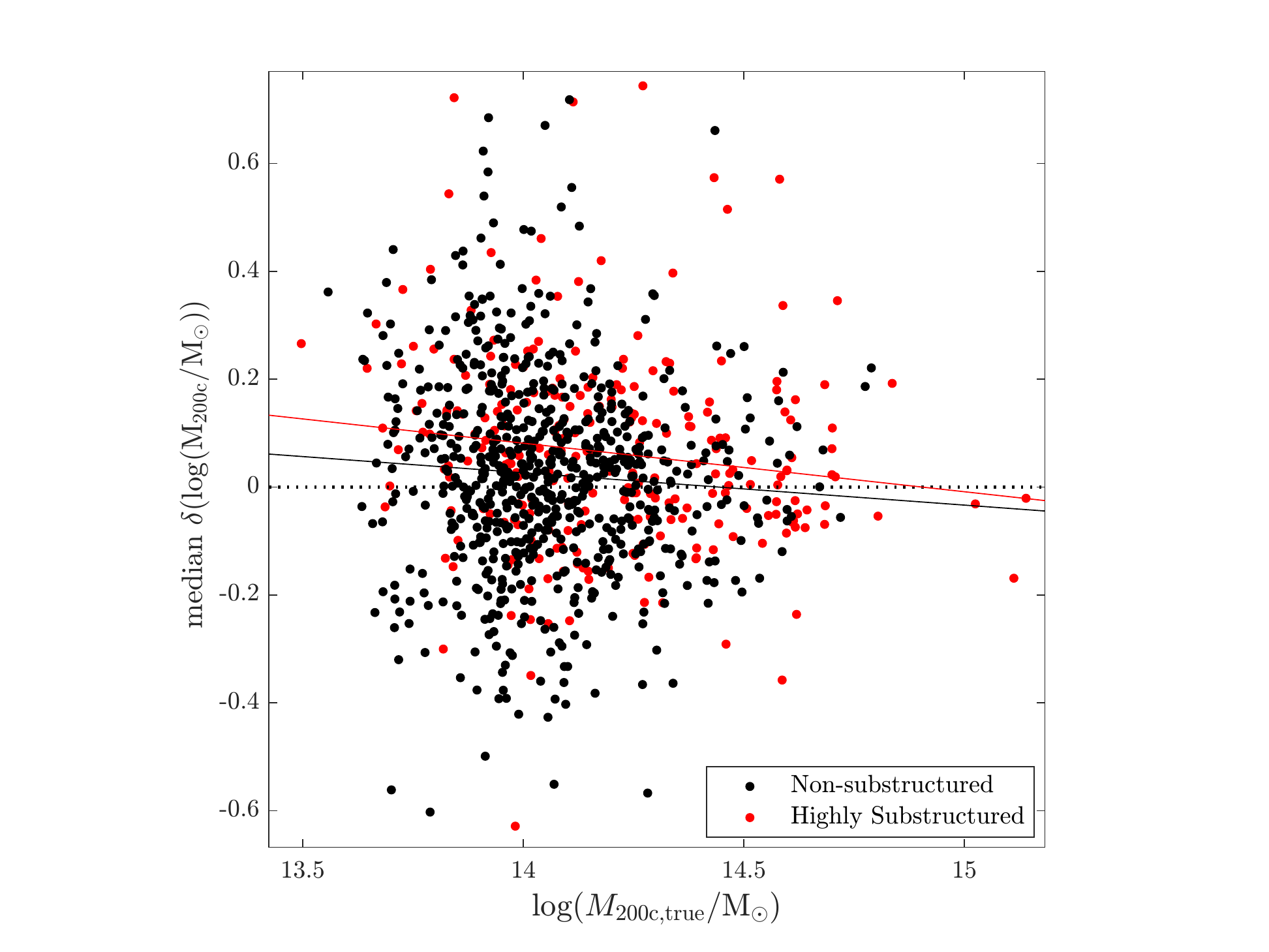}
 \caption{The median difference between recovered and true log mass versus true log mass, $\delta_{\rm M200c} = \rm{log}(M_{\rm 200c, rec})-\rm{log}(M_{\rm 200c, true})$ for all 23 methods. Clusters that deemed as highly dynamically substructured by the DS or Kappa test are denoted as red circles, and the non-substructured clusters denoted are by black circles. The red line reflects a linear fit for the substructured clusters of $\delta_{\rm M200c} = -0.06\;(\rm{log}(M_{\rm 200c, rec})-14.126)+0.019$. The black line reflects a linear fit for the non-substructured clusters of $\delta_{\rm M200c} = -0.09\;(\rm{log}(M_{\rm 200c, rec})-14.126)+0.07$.}
\label{fig:SAM2_SW_median_delta_M_Mtrue_clusters}
\end{figure}

\label{lastpage}
\end{document}